\documentclass[a4paper, 10pt, conference]{IEEEtran}
\IEEEoverridecommandlockouts
\usepackage{cite}
\usepackage{amsmath,amssymb,amsfonts}
\usepackage{graphicx}
\usepackage{textcomp}
\usepackage{adjustbox}
\usepackage{balance}
\usepackage{threeparttable}
\usepackage{soul}
\usepackage{verbatim}
\usepackage{url}
\usepackage{amssymb}
\usepackage{mathtools}
\usepackage{xparse}
\usepackage[percent]{overpic}
\usepackage{textcomp}
\usepackage{comment}
\usepackage{multirow}
\usepackage{array}
\usepackage{nicefrac} 
\usepackage{algorithm} 
\usepackage{algpseudocode} 
\usepackage[hidelinks]{hyperref}
\usepackage{siunitx, booktabs}
\hypersetup{
    colorlinks=true,
    linkcolor=black,
    filecolor=none,      
    urlcolor=blue,
    citecolor=black,
    pdftitle={Overleaf Example},
    pdfpagemode=FullScreen,
    }
\usepackage[table, svgnames, dvipsnames]{xcolor}

\AtBeginEnvironment{matrix}{\setlength{\arraycolsep}{0.45ex}}

\makeatletter
\newcommand{\raisemath}[1]{\mathpalette{\raiseMath{#1}}}%
\newcommand{\raiseMath}[3]{\raisebox{#1}[0pt][0pt]{$#2#3$}}

\makeatother

\NewDocumentCommand{\qbar}{O{0.5pt} O{-6.55pt}}{
	\ensuremath{\mathrlap{\raisemath{#2}{\hspace*{#1}{\mathchar'26\mkern-9mu}}} q}%
}

\NewDocumentCommand{\qbarsmall}{O{0.5pt} O{-6.0pt}}{
	\ensuremath{\mathrlap{\raisemath{#2}{\hspace*{#1}{\mathchar'26\mkern-9mu}}} q}%
}

\NewDocumentCommand{\qbarssmall}{O{0.5pt} O{-4.0pt}}{
	\ensuremath{\mathrlap{\raisemath{#2}{\hspace*{#1}{\mathchar'26\mkern-9mu}}} q}%
}

\NewDocumentCommand{\qbars}{O{0.5pt} O{-4.65pt}}{
	\ensuremath{\mathrlap{\raisemath{#2}{\hspace*{#1}{\mathchar'26\mkern-9mu}}} q}%
}

\NewDocumentCommand{\qbarc}{O{0.5pt} O{-5.2pt}}{
	\ensuremath{\mathrlap{\raisemath{#2}{\hspace*{#1}{\mathchar'26\mkern-9mu}}} q}%
}

\NewDocumentCommand{\pbar}{O{-1.5pt} O{-6.65pt}}{
	\ensuremath{\mathrlap{\raisemath{#2}{\hspace*{#1}{\mathchar'26\mkern-9mu}}} p}%
}

\newcommand{\bs}[1]{\boldsymbol{#1}}  
\newcommand{\ts}[1]{\text{#1}}

\renewcommand{\baselinestretch}{1.0}  

\def\BibTeX{{\rm B\kern-.05em{\sc i\kern-.025em b}\kern-.08em
    T\kern-.1667em\lower.7ex\hbox{E}\kern-.125emX}}
\begin{document}

\title{\LARGE \bf \mbox{Closed-Loop} Stability of a \mbox{Lyapunov-Based} Switching Attitude Controller for \mbox{Energy-Efficient} \mbox{Torque-Input-Selection} During Flight\\
\thanks{This work was supported by the Joint Center for Aerospace Technology Innovation (JCATI) through \mbox{Award\,$172$}, the Washington State University (WSU) Foundation and the Palouse Club through a Cougar Cage Award to \mbox{N.\,O.\,P\'erez-Arancibia}, and the WSU Voiland College of Engineering and Architecture through a \mbox{start-up} fund to \mbox{N.\,O.\,P\'erez-Arancibia}.}
\thanks{\mbox{F.\,M.\,F.\,R.\,Gon\c{c}alves} and \mbox{N.\,O.\,P\'erez-Arancibia} are with the School of Mechanical and Materials Engineering, Washington State University (WSU), Pullman, \mbox{WA~$99164$-$2920$}, USA. \mbox{R.\,M.\,Bena} is with the Department of Mechanical and Civil Engineering, California Institute of Technology (Caltech), Pasadena, \mbox{CA~$91125$-$2100$}, USA. Corresponding authors' Email:~{\tt francisco.goncalves@wsu.edu}~(\mbox{F.\,M.\,F.\,R.\,G.}); {\tt n.perezarancibia@wsu.edu}~(\mbox{N.\,O.\,P.-A.})}
}

\author{\mbox{Francisco M. F. R. Gon\c{c}alves}, Ryan M. Bena, and \mbox{N\'{e}stor O. P\'{e}rez-Arancibia}}

\maketitle
\thispagestyle{empty}
\pagestyle{empty}

\begin{abstract}
We present a new \mbox{Lyapunov-based} switching attitude controller for \mbox{energy-efficient} \mbox{real-time} selection of the torque inputted to an \textit{uncrewed aerial vehicle} (UAV) during flight. The proposed method, using quaternions to describe the attitude of the controlled UAV, interchanges the stability properties of the two fixed points---one locally asymptotically stable and another unstable---of the resulting \textit{\mbox{closed-loop}} (CL) switching dynamics of the system. In this approach, the switching events are triggered by the value of a compound \mbox{energy-based} function. To analyze and ensure the stability of the CL switching dynamics, we use classical nonlinear Lyapunov techniques, in combination with \mbox{switching-systems} theory. For this purpose, we introduce a new compound \textit{Lyapunov function} (LF) that not only enables us to derive the conditions for CL asymptotic and exponential stability, but also provides us with an estimate of the CL system's region of attraction. This new estimate is considerably larger than those previously reported for systems of the type considered in this paper. To test and demonstrate the functionality, suitability, and performance of the proposed method, we present and discuss experimental data obtained using a \mbox{$\bs{31}$-g} quadrotor during the execution of \mbox{high-speed} \mbox{yaw-tracking} maneuvers. Also, we provide empirical evidence indicating that all the initial conditions chosen for these maneuvers, as estimated, lie inside the system's region of attraction. Last, experimental data obtained through these flight tests show that the proposed switching controller reduces the control effort by about \mbox{$\bs{53}$\,\%}, on average, with respect to that corresponding to a \mbox{commonly~used} benchmark control scheme, when executing a particular type of \mbox{high-speed} \mbox{yaw-tracking} maneuvers.
\end{abstract}

\section{Introduction}
\label{Section01}
We envision a future in which swarms of robotic flying insects are deployed in unstructured environments to perform tasks useful for society; for example, biological research that requires the physical and visual tracking of natural insects. The execution of these types of maneuvers necessitates the use of \mbox{high-performance}, robust, and efficient controllers capable of following \mbox{high-speed} trajectories while maintaining stable flight. We can imagine, for instance, a robotic insect flying embedded in a bee colony; then, while tracking a particular specimen the robot is commanded to track a different bee and must promptly rotate in order to follow the new reference. Intuitively, it seems that we should always apply the control torque in the direction of the shorter rotational path; however, depending on the instantaneous attitude and \mbox{angular-velocity} errors of the flyer, it might be advantageous to apply the control torque in the direction of the longer rotational path, according to a \mbox{user-defined}~\textit{performance figure of merit} (PFM)~\cite{Schlanbusch2010Choosing,Mayhew_Robust,QuatAutomatica, Goncalves2024ICRA, Goncalves2024ROBOT}. 

In this paper, we use quaternions to represent the attitude kinematics and dynamics of the controlled \textit{uncrewed aerial vehicle} (UAV) in space, a method widely used in aerial robotics and spacecraft control due to its numerical robustness and suitability to avoid issues caused by kinematic \mbox{singularities~\cite{Goncalves2024ICRA,Goncalves2024ROBOT,BenaMPPC2022,Bena2023Yaw,calderon2019control,yang2019bee, Salcudean, Wie_Sign, Thienel_sign, Kristiansen,Fjellstad_sign,Schlanbusch2010Choosing,Bena_RAL_Perception,bhat2000topological,Mayhew_Robust,Ying_ICRA_2016, Ying_IEEE_TCST_2020, Ying_ACC_2017, Ying_IROS_2018, Ying_ICRA_2019, Ying_Automatica_2024,QuatAutomatica,comparisonEulerQuaternion}}. Unfortunately, as discussed in~\cite{Goncalves2024ICRA,Goncalves2024ROBOT,BenaMPPC2022,Bena2023Yaw}, when quaternions are used to define attitude controllers of the type presented in~\cite{Bena2023Yaw}, the resulting \textit{\mbox{closed-loop}} (CL) dynamics exhibit two fixed points corresponding to the same attitude kinematics but with opposite stability properties---one is locally asymptotically stable and the other is unstable. This issue brings challenges regarding flight performance; however, the stability properties of the two CL system's fixed points can be interchanged in real time by simply changing the sign of a term in the control law---equivalent to reversing the direction of the torque input corresponding to that term---which is beneficial from a performance perspective in some cases. Specifically, a switching scheme of this type can be implemented to prevent unwinding behavior, defined as large rotations (\mbox{$>\pi$\,rad}) that return the flyer to its original orientation and are caused by representational ambiguities~\cite{bhat2000topological}. The most common technique used to avoid unwinding is to multiply the term \textit{proportional} to the vector part of the \textit{\mbox{attitude-error} quaternion} (AEQ) by the sign of its scalar part in the definition of the control law that specifies the total torque inputted to the UAV's \mbox{open-loop} \mbox{dynamics~\cite{Salcudean, Wie_Sign, Thienel_sign, Kristiansen, Fjellstad_sign, Bena_RAL_Perception, Bena2023Yaw, BenaMPPC2022, yang2019bee, calderon2019control}}. This scheme ensures that the direction of the proportional torque is aligned with that of the shorter rotational path, which is not always the best decision from an energy perspective. 

To address this \mbox{energy-motivated} decision problem, \cite{Schlanbusch2010Choosing} presents a heuristic method based on the \mbox{\textit{a-priori}} specification of a set of rules extracted from \mbox{one-thousand} simulations with random initial conditions for orientation and angular velocity; \cite{Mayhew_Robust} presents two hybrid controllers based on backstepping and the definition of an \mbox{energy-like} \textit{Lyapunov function} (LF), respectively; \cite{QuatAutomatica} presents a hybrid controller that accounts for both the AEQ and \mbox{angular-velocity} error in the definition of a switching law; and, \cite{Goncalves2024ICRA} presents a \mbox{model-predictive} method to select the most \mbox{cost-efficient} direction of the proportional torque input according to a \mbox{user-defined} PFM. More recently, in~\cite{Goncalves2024ROBOT}, we introduced a \mbox{Lyapunov-based} switching attitude controller that accounts for both the AEQ and the \mbox{angular-velocity} error to select, between two options, the torque law used for feedback control during flight; however, the chosen LF allows only for a very conservative estimation of the region of attraction of the stable CL fixed state, and we did not show that the unstable equilibrium is a saddle point---a topic thoroughly discussed in this paper. Here, we present a new \mbox{Lyapunov-based} switching attitude controller that selects the torque inputted to the controlled UAV, according to a criterion of energy efficiency, in real time. Also, we present stability analyses of the CL dynamics resulting from using the proposed switching scheme. Furthermore, by using a modified version of the LF in~\cite{Goncalves2024ROBOT}, we show that one of the two fixed states of the CL system is exponentially stable under a set of conditions, and provide an estimate of the corresponding region of attraction. To test the proposed approach, we implemented the new switching controller on a \mbox{$31$-g} quadrotor to execute \mbox{high-speed} \mbox{yaw-tracking} maneuvers. The obtained experimental data compellingly demonstrate the suitability, functionality, and performance of the proposed approach.
\begin{figure}[t!]
\vspace{1ex}
\begin{center}
\includegraphics{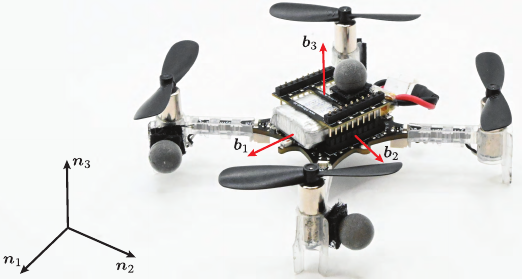}
\end{center}
\vspace{-3ex}     
\caption{\textbf{UAV platform used in the \mbox{real-time} flight experiments, the \mbox{Crazyflie\,$\bs{2.1}$}.} Here, \mbox{$\bs{\mathcal{B}} = \left\{\bs{b}_1,\bs{b}_2,\bs{b}_3 \right\}$}, with its origin coinciding with the UAV's center of mass, denotes the \mbox{body-fixed} frame of reference; \mbox{$\bs{\mathcal{N}} = \left\{\bs{n}_1,\bs{n}_2,\bs{n}_3 \right\}$} denotes the inertial frame of reference. \label{FIG01}} 
\vspace{-2ex}
\end{figure}

The rest of the paper is organized as follows. \mbox{Section\,\ref{Section02}} reviews the main topics regarding the attitude dynamics of the controlled UAV and briefly describes a \mbox{commonly-used} \mbox{quaternion-based} continuous controller that we employed as the starting point of the presented research. \mbox{Section\,\ref{Section03}} discusses the derivation of the two fixed points of the CL system resulting from using the continuous controller and their stability properties; this section also formulates and explains the performance problem associated with the implementation of \mbox{quaternion-based} attitude control laws of the type considered here. \mbox{Section\,\ref{Section04}} describes the switching control scheme introduced in this paper, derives the two equilibrium points---one stable and another unstable---of the resulting CL system, analyzes their stability, and provides an estimate for the region of attraction of the stable CL fixed point. \mbox{Section\,\ref{Section05}} presents and analyzes experimental results. Last, \mbox{Section\,\ref{Section06}} states some conclusions.

\vspace{2ex}
\textit{\textbf{Notation:}}
\begin{enumerate}
\item\,$\mathbb{R}$ and $\mathbb{R}^3$ denote the sets of real numbers and triplets, respectively.
\item\,$\mathcal{S}^3$ denotes the set of unit quaternions.
\item\,Italic lowercase symbols denote scalars, e.g., $p$; bold lowercase symbols denote vectors, e.g., $\bs{p}$; bold uppercase symbols denote matrices, e.g., $\bs{P}$; and bold crossed lowercase symbols denote quaternions, \mbox{e.g., $\bs{\pbar}$}.
\item\,The symbols $\times$ and $\otimes$ denote the vector \mbox{cross-product} and quaternion product, respectively.
\item The operator \mbox{$\| \cdot \|_2$} computes the $2$-norm of a vector.
\item The operator \mbox{$\ts{sgn} \left\{\,\cdot\,\right\}$} extracts the sign of a real scalar.
\item The symbols $>$, $<$, $\geq$, and $\leq$ denote ordering or definiteness relationships when used with scalars or matrices, respectively.
\item The symbol $\bs{I}$ denotes any identity matrix of adequate dimensions; the symbol $\bs{0}$ denotes any block of zeros of adequate dimensions.
\end{enumerate}

\section{Attitude Dynamic Model of the UAV and Continuous Controller}
\label{Section02}
\subsection{Rigid-Body Attitude Dynamics}
\label{Subsection02A}
Here, we briefly review the attitude dynamics of the controlled UAV used in the experiments, which is shown in~\mbox{Fig.\,\ref{FIG01}}; a description of the translational dynamics can be found in~\cite{BenaMPPC2022}. \mbox{Fig.\,\ref{FIG01}} also graphically defines the frames used to represent the kinematics of the flyer. Namely, the \mbox{body-fixed} frame, \mbox{$\bs{\mathcal{B}} = \left\{\bs{b}_1,\bs{b}_2,\bs{b}_3 \right\}$}, with its origin coinciding with the UAV's center of mass; and, the inertial frame, \mbox{$\bs{\mathcal{N}} = \left\{\bs{n}_1,\bs{n}_2,\bs{n}_3 \right\}$}. As discussed in\mbox{~\cite{BenaMPPC2022, Bena2023Yaw, calderon2019control,yang2019bee}}, using quaternions to describe the UAV's attitude in space and D'Alembert's law, the \mbox{open-loop} dynamics of the system evolve according to
\begin{align}
\begin{split}
\bs{\dot{\qbar}} &= \frac{1}{2} \bs{\qbar}\otimes
\begin{bmatrix}
0 \\
\bs{\omega} 
\end{bmatrix},\\
\bs{\dot{\omega}} &=  \bs{J}^{-1}\left(\bs{\tau}-\bs{\omega}\times \bs{J}\bs{\omega}\right),
\label{EQ01}
\end{split}
\end{align}
in which $\bs{\qbar}$ is a unit quaternion that represents the orientation of $\bs{\mathcal{B}}$ relative to $\bs{\mathcal{N}}$; $\bs{\omega}$ is the angular velocity of $\bs{\mathcal{B}}$ with respect to $\bs{\mathcal{N}}$, written in $\bs{\mathcal{B}}$ coordinates; $\bs{J}$ is the inertia matrix of the UAV, written in $\bs{\mathcal{B}}$ coordinates; and, $\bs{\tau}$ is the torque inputted to the system and, in closed loop, generated by a control law. Note that \mbox{$\bs{\qbar} = \left[m\,\,\bs{n}\right]^T$}, with \mbox{$m = \cos{\frac{\Phi}{2}}$} and \mbox{$\bs{n} = \bs{u}\sin{\frac{\Phi}{2}}$}, where $\Phi$ is the amount that $\bs{\mathcal{N}}$ must be rotated about $\bs{u}$ to become exactly aligned with $\bs{\mathcal{B}}$.

\subsection{Continuous Feedback \mbox{Quaternion-Based} Controller}
\label{Subsection02B}
Now, we briefly review the continuous controller used as the starting point of the main research presented in this paper. As customary, we define the attitude control error using the AEQ, i.e.,  
\begin{align}
\bs{\qbar}_\text{e} = \bs{\qbar}^{-1} \otimes \bs{\qbar}_\text{d},
\label{EQ02}
\end{align}
in which $\bs{\qbar}_{\ts{d}}$ is the desired attitude quaternion corresponding to the orientation of the desired \mbox{body-fixed} frame, $\bs{\mathcal{B}}_{\ts{d}}$, with respect to $\bs{\mathcal{N}}$. Also, as explained in~\cite{BenaMPPC2022}, \mbox{$\bs{\qbar}_{\ts{e}} = \left[m_{\ts{e}}\,\,\bs{n}_{\ts{e}}\right]^T$} represents the orientation of $\bs{\mathcal{B}}_{\ts{d}}$ relative to $\bs{\mathcal{B}}$; namely, \mbox{$m_{\ts{e}} = \cos{\frac{\Phi_{\ts{e}}}{2}}$} and \mbox{$\bs{n}_{\ts{e}} = \bs{u}_{\ts{e}}\sin{\frac{\Phi_{\ts{e}}}{2}}$}, where $\Phi_{\ts{e}}$ is the amount that $\bs{\mathcal{B}}$ must be rotated about $\bs{u}_{\ts{e}}$ to become perfectly aligned with $\bs{\mathcal{B}}_{\ts{d}}$. In this case, the control law used to compute the torque inputted to the system is
\begin{align}
\bs{\tau} =  \bs{J}\left( k_{\bs{\qbars}}\bs{n}_{\ts{e}} + k_{\bs{\omega}} \bs{\omega}_{\ts{e}} + \bs{\dot{\omega}}_\ts{d}\right) + \bs{\omega}\times\bs{J}\bs{\omega},
\label{EQ03}
\end{align}
in which the controller gains $k_{\bs{\qbars}}$ and $k_{\bs{\omega}}$ are positive scalars; $\bs{\omega}_{\ts{d}}$ is the desired angular velocity, chosen to be kinematically consistent with $\bs{\qbar}_{\ts{d}}$; and, \mbox{$\bs{\omega}_{\ts{e}} = \bs{\omega}_{\ts{d}} - \bs{\omega}$} is the \mbox{angular-velocity} error. This structure corresponds to a \mbox{proportional-derivative} controller (first two terms) with additional feedforward and \mbox{feedback-linearization} control actions (last two terms). 

\section{Attitude Dynamics and Performance Problem}
\label{Section03}
\subsection{Equilibrium Points of the CL Attitude System}
\label{Subsection03A}
To obtain the \mbox{state-space} representation of the resulting CL dynamics, we differentiate~(\ref{EQ02}) with respect to time and plug the right side of~(\ref{EQ03}) into the \mbox{open-loop} dynamics equation specified by~(\ref{EQ01}), which yields
\begin{align}
\begin{split}
\bs{\dot{\qbar}}_{\ts{e}} &= \frac{1}{2}
\left[
\begin{array}{c}
    0 \\
    \bs{\omega}_{\ts{e}}
\end{array}
\right]
\otimes \bs{\qbar}_{\text{e}},\\
    \bs{\dot{\omega}}_{\ts{e}} 
    &= -\left(k_{\bs{\qbars}} \bs{n}_{\ts{e}} + k_{\bs{\omega}} \bs{\omega}_{\ts{e}}\right).
\end{split}
\label{EQ04}
\end{align}
As explained in~\cite{BenaMPPC2022}, this system has two fixed points; one was proven to be locally asymptotically stable and the other unstable. These two equilibria can be found by simply setting \mbox{$\bs{\dot{\qbar}}_{\ts{e}} = \left[ 0\,\, 0\,\, 0\,\, 0 \right]^T$} and \mbox{$\bs{\dot{\omega}}_{\ts{e}} = \left[0\,\, 0\,\, 0 \right]^T$}. The stable equilibrium point corresponds to the pair \mbox{$\bs{\qbar}_{\ts{e}}^{\star} = \left[+1\,\, 0\,\, 0\,\, 0 \right]^T$}~and~\mbox{$\bs{\omega}_{\ts{e}}^{\star} =  \left[ 0\,\, 0\,\, 0 \right]^T$}; the unstable equilibrium point corresponds to the pair~\mbox{$\bs{\qbar}_{\ts{e}}^{\dagger} = \left[-1\,\, 0\,\, 0\,\, 0 \right]^T$}~and~$\bs{\omega}_{\ts{e}}^{\star}$. Even though these two fixed points have opposite stability properties, they represent the same kinematic configuration; i.e., identical attitude and \mbox{angular-velocity} errors. This issue is connected to the \mbox{double-covering-of-rotations} phenomenon discussed in~\cite{bhat2000topological} and caused by \textit{\mbox{quaternion-sign} ambiguity}. Notably, their stability properties can be interchanged by simply making $k_{\bs{\qbars}}$ negative.

\subsection{Performance Problem}
\label{Subsection03B}
As discussed in~\mbox{Section\,\ref{Section01}}, the control law specified by~(\ref{EQ03}) induces large unwanted rotations (\mbox{$>\pi$\,rad}) in some situations~\cite{Goncalves2024ICRA}. This dynamical behavior can be modulated in real time by switching the sign of \mbox{$k_{\bs{\qbars}}$} and, as a consequence, swapping the stability properties---including the region of attraction---of the two fixed points of the CL system. By changing the selection of one control law over the other---determined by choosing \mbox{$k_{\bs{\qbars}} > 0$} or \mbox{$k_{\bs{\qbars}} < 0$}---the CL system is compelled to converge to a different equilibrium point and, therefore, to dynamically behave differently during operation. If this dynamical selection is made intelligently, according to a desired objective, it is possible to improve control performance. The most common method used in the past to implement a simple switching control law of this type is to multiply the first term in~(\ref{EQ03}) by the signum function of $m_{\ts{e}}$. Namely,
\begin{align}
\bs{\tau}_{\ts{b}} = \bs{J}\left( \ts{sgn}\{m_\ts{e}\}k_{\bs{\qbars}}\bs{n}_{\ts{e}} + k_{\bs{\omega}} \bs{\omega}_{\ts{e}} + \bs{\dot{\omega}}_\ts{d}\right) + \bs{\omega}\times\bs{J}\bs{\omega}.
\label{EQ05}
\end{align}
This modification changes the resulting CL dynamics in two fundamental ways. First, when the sign of $m_{\ts{e}}$ changes, the stability properties of the two CL equilibria are reversed. Second, the torque component $\ts{sgn}\{m_\ts{e}\}k_{\bs{\qbars}}\bs{n}_{\ts{e}}$ is always applied in the direction of the shorter rotational path. This simple switching scheme is profusely used in the fields of aerial robotics and spacecraft control, and has been empirically demonstrated to be very effective; however, it does not account for the \mbox{angular-velocity} error and, therefore, does not guarantee the best performance in terms of energy~\cite{Goncalves2024ICRA,Goncalves2024ROBOT}.

\section{A New Switching Attitude Controller}
\label{Section04}
\subsection{Definition of the Torque Control Law}
\label{Section04A}
To address the issue discussed in \mbox{Section\,\ref{Subsection03B}}, we define a new switching attitude control law as
\begin{align}
\begin{split}
\hspace{-2ex} \bs{\tau}_{\hspace{-0.2ex}\sigma} = \bs{J}\left(\sigma k_{\bs{\qbarssmall}}\bs{n}_{\ts{e},\sigma}  + k_{\bs{\omega}} \bs{\nu}_{\sigma} + \bs{\dot{\omega}}_{\ts{d}} + \sigma k_{\bs{n}} \bs{\dot{n}}_{\ts{e},\sigma} \right) + \bs{\omega}\times\bs{J}\bs{\omega},
\end{split}
\label{EQ06}
\end{align}
in which the variable \mbox{$\sigma \in \left\{+1, -1\right\}$} is the switching signal; \mbox{$\bs{\nu}_{\sigma}= \bs{\omega}_{\ts{e}} + \sigma k_{\bs{n}}\bs{n}_{\ts{e},\sigma}$}; and, \mbox{$0 < k_{\bs{n}} \in \mathbb{R}$}. Note that what makes (\ref{EQ06}) different from (\ref{EQ03}) and (\ref{EQ05}) is the inclusion of $\bs{\dot{n}}_{\ts{e},\sigma}$ in its definition, which enables us to specify the switching function $\Lambda$---which selects $\sigma$ in real time---as a function of both the AEQ and the \mbox{angular-velocity} error. This selection process is further explained in \mbox{Section\,\ref{SubSection04C}}. For the \mbox{real-time} implementation of (\ref{EQ06}), we define the switching condition as
\begin{align}
\sigma^{+} =
\left\{
\begin{array}{ccc}
~~\sigma, & \ts{if} & -\delta < \Lambda < \delta \\
+1, & \ts{if} & ~~~~~~~\,\Lambda \geq \delta   \\
-1, & \ts{if} & ~~~~~~~~~\,\Lambda \leq -\delta 
\end{array}
\right.,
\label{EQ07}
\end{align}
where $\sigma^{+}$ is the value of the switching signal to be applied the next time instant; $\sigma$ is the current value of the switching signal, $+1$ or $-1$; and, \mbox{$0<\delta\in\mathbb{R}$} is the hysteresis margin.

\subsection{Equilibrium Points of the New CL System} 
\label{Subsection04B}
As shown in~\cite{Goncalves2024ROBOT}, the new CL system resulting from using the control input specified by (\ref{EQ06}) and (\ref{EQ07}) can be found by substituting (\ref{EQ06}) into the last row of (\ref{EQ01}), which yields
\begin{align}
\begin{split}
\bs{\dot{\qbar}}_{\ts{e},\sigma} &= \frac{1}{2}
\left[
\begin{array}{c}
    0 \\
    \bs{\nu}_{\sigma} - \sigma k_{\bs{n}}\bs{n}_{\ts{e},\sigma}
\end{array}
\right]
\otimes \bs{\qbar}_{\text{e},\sigma},\\
    \bs{\dot{\nu}}_{\sigma} 
    &= -\left( \sigma k_{\bs{\qbars}} \bs{n}_{\ts{e},\sigma} + k_{\bs{\omega}} \bs{\nu}_{\sigma} \right). \\
\end{split}
\label{EQ08}
\end{align}
Next, by setting \mbox{$\bs{\dot{\qbar}}_{\ts{e},\sigma} = \left[ 0\,\, 0\,\, 0\,\, 0 \right]^T$} and \mbox{$\bs{\dot{\nu}}_{\sigma} = \left[0\,\, 0\,\, 0 \right]^T$}, and then solving the resulting set of algebraic equations, it can be verified that each subsystem has two fixed points. For both subsystems, the first equilibrium corresponds to the pair \mbox{$\bs{\qbar}_{\ts{e},\sigma}^{\star} = \left[ +1\,\, 0\,\, 0\,\, 0 \right]^T$} and \mbox{$\bs{\nu}_{\sigma}^{\star} =  \left[ 0\,\, 0\,\, 0 \right]^T$}, 
and the second equilibrium corresponds to the pair \mbox{$\bs{\qbar}_{\ts{e},\sigma}^{\dagger} = [-1\,\, 0\,\, 0\,\, 0]^T$} and \mbox{$\bs{\nu}_{\sigma}^{\star}$}, with \mbox{$\sigma \in \left\{+1, -1\right\}$}.
  
\subsection{Stability Analysis of the Switching Dynamics}
\label{SubSection04C}
To analyze and enforce the stability of the CL switching system specified by~(\ref{EQ08}), we use \mbox{Theorem\,$3.1$} in~\cite{liberzon2003switching} and \textit{Lyapunov's direct method} as stated in Theorem\,$4.9$ of~\cite{khalil2002nonlinear}. The following proposition formalizes the idea.
\vspace{1ex}

\noindent\textbf{Proposition\,$\bs{1}$.} \textit{Let the references} $\bs{\qbar}_\ts{d}$ \textit{and} $\bs{\omega}_\ts{d}$ \textit{be smooth and bounded functions of time, and let} $k_{\bs{\qbars}}$, $k_{\bs{\omega}}$, \textit{and} $k_{\bs{n}}$ \textit{be constant real positive scalars. Then, the fixed set} \mbox{$\{\sigma \bs{\qbar}^{\star}_\ts{e}, \bs{\nu}^{\star}_{\sigma}\}$}, \textit{for} \mbox{$\sigma \in \left\{+1,-1\right\}$}, \textit{of the CL \mbox{state-space} switching attitude dynamics specified by (\ref{EQ08}),} \textit{with} \mbox{$\bs{\qbar}_{\ts{e}}^{\star} = \left[ +1\,\, 0\,\, 0\,\, 0 \right]^T\hspace{-0.4ex}$} \textit{and} \mbox{$\bs{\nu}_{\sigma}^{\star} = \left[ 0\,\, 0\,\, 0 \right]^T\hspace{-0.4ex}$}, \textit{is locally asymptotically stable.}
\vspace{1ex}

\noindent\textit{Proof.}~
First, for the subsystem corresponding to \mbox{$\sigma = +1$}, we show that the equilibrium point \mbox{$\{\bs{\qbar}^{\star}_{\ts{e},+1}, \bs{\nu}^{\star}_{+1} \}$} is locally asymptotically stable. With this purpose, we define
\begin{align}
\begin{split}
V_{+1}(\bs{\qbar}_{\ts{e}}, \bs{\nu}_{+1}) = \frac{1}{2} k_{\bs{\qbars}}^{-1} \bs{\nu}_{+1}^T \bs{\nu}_{+1} + 2c(1 - m_\ts{e}),
\end{split}
\label{EQ09}
\end{align}
with \mbox{$0<c\in \mathbb{R}$}, which can be proven to be an LF for the CL system specified by (\ref{EQ08}). This fact can be readily concluded by noticing that 
\begin{align}
\begin{split}
V_{+1}(\bs{\qbar}_{\ts{e}},\bs{\nu}_{+1}) &> 0,~~\forall~\left\{ \bs{\qbar}_{\ts{e}},\bs{\nu}_{+1} \right\} \neq \left\{ \bs{\qbar}_{\ts{e}}^{\star},\bs{\nu}_{+1}^{\star} \right\}, \\
V_{+1}(\bs{\qbar}^{\star}_{\ts{e}}, \bs{\nu}^{\star}_{+1}) &= 0,
\end{split}
\label{EQ10}
\end{align}
because \mbox{$k_{\bs{\qbars}} > 0$}, and that the function \mbox{$\dot{V}_{+1}(\bs{\qbar}_{\ts{e}},\bs{\nu}_{+1}) < 0$}, for all \mbox{$\left\{ \bs{\qbar}_{\ts{e}},\bs{\nu}_{+1} \right\} \not\in \left\{ \bs{\qbar}_{\ts{e}}^{\star},\bs{\nu}_{+1}^{\star} \right\}
\cup
\{ \bs{\qbar}_{\ts{e}}^{\dagger},\bs{\nu}_{+1}^{\star} \}$}, with properly chosen values for $c$, $k_{\bs{n}}$, $k_{\bs{\omega}}$, and $k_{\bs{\qbars}}$. We present the mathematical argument for this last statement next.

The time derivative of $V_{+1}(\bs{\qbar}_{\ts{e}},\bs{\nu}_{+1})$  as specified by~(\ref{EQ09}) is
\begin{align}
\dot{V}_{+1}(\bs{\qbar}_{\ts{e}},\bs{\nu}_{+1}) = 
&-\bs{\nu}^T_{+1}\bs{n}_\ts{e} -k_{\bs{\qbars}}^{-1}k_{\bs{\omega}} \bs{\nu}_{+1}^T \bs{\nu}_{+1} - 2c\dot{m}_{\ts{e}}.
\label{EQ11}
\end{align}
Furthermore, for \mbox{$\sigma = +1$}, it can be shown that
\begin{align}
\dot{m}_{\ts{e}} = -\frac{1}{2}\bs{\nu}_{+1}^T\bs{n}_{\ts{e}} + \frac{1}{2} k_{\bs{n}} \bs{n}^T_\ts{e}\bs{n}_\ts{e},
\label{EQ12}
\end{align}
which allows us to rearrange and \mbox{upper-bound}~(\ref{EQ11}) as
\begin{align}
\begin{split}
\dot{V}_{+1}(\bs{\qbar}_{\ts{e}},\bs{\nu}_{+1}) &= (c-1)\bs{\nu}_{+1}^T\bs{n}_{\ts{e}} -k_{\bs{\qbars}}^{-1}k_{\bs{\omega}} \|\bs{\nu}_{+1}\|_2^2 - ck_{\bs{n}} \|\bs{n}_\ts{e}\|_2^2\\
&\leq c\|\bs{\nu}_{+1}\|_2\|\bs{n}_{\ts{e}}\|_2 -k_{\bs{\qbars}}^{-1}k_{\bs{\omega}} \|\bs{\nu}_{+1}\|_2^2 - ck_{\bs{n}} \|\bs{n}_\ts{e}\|_2^2\\
&= -\bs{x}^T \bs{P} \bs{x},
\end{split}
\label{EQ13}
\end{align}
where \mbox{$\bs{x} = \left[\|\bs{n}_\ts{e}\|_2\,\,\|\bs{\nu}_{+1}\|_2\right]^T$}  and 

\begin{align}
\begin{split}
\bs{P} =
\left[
\hspace{-0.5ex}
\begin{array}{cc}
ck_{\bs{n}} & -\frac{c}{2}\\[1ex]
 -\frac{c}{2} & k_{\bs{\qbars}}^{-1}k_{\bs{\omega}}
\end{array}
\hspace{-0.5ex}
\right]
> 0,
\end{split}
\end{align}
for a sufficiently small value of $c$; namely, \mbox{$c < 4k_{\bs{n}}k_{\bs{\omega}}k^{-1}_{\bs{\qbars}}$}. Therefore, since $V_{+1}(\bs{\qbar}_{\ts{e}}, \bs{\nu}_{+1})$ is an LF for the system specified by (\ref{EQ08}), we conclude that the equilibrium corresponding to the pair \mbox{$\{\bs{\qbar}^{\star}_{\ts{e},+1}, \bs{\nu}^{\star}_{+1} \}$} is locally asymptotically stable.
\begin{figure*}[t!]
\vspace{1ex}
\begin{center}
\includegraphics{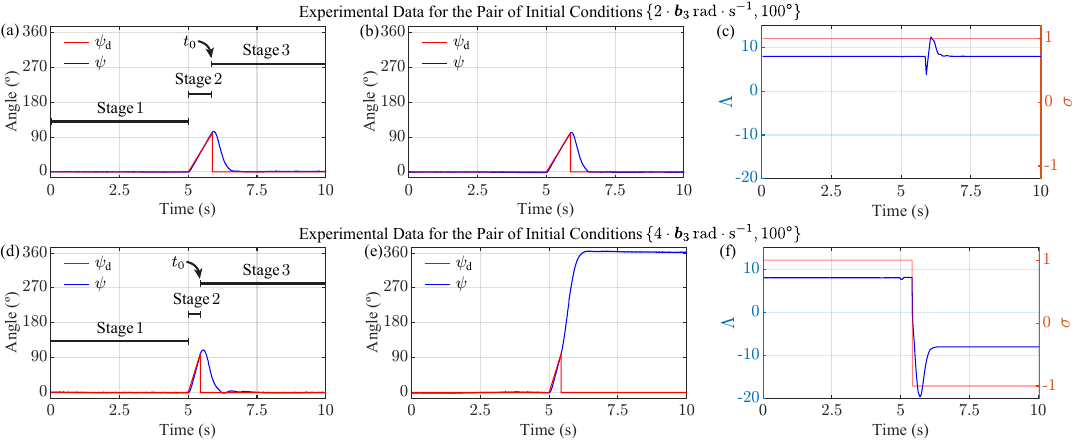}
\end{center}
\vspace{-3ex}
\caption{\textbf{Experimental data corresponding to two sets of flight tests with different pairs of initial conditions}. \textbf{(a)}~Reference and measured yaw angle, for the pair of initial conditions \mbox{$\left\{ \bs{\omega}_0, \psi_0 \right\} = \{2\cdot\bs{b}_3\,\ts{rad}\cdot\ts{s}^{-1},100^\circ\}$}, obtained using the benchmark controller. \textbf{(b)}~Reference and measured yaw angle, for the pair of initial conditions \mbox{$\left\{ \bs{\omega}_0, \psi_0 \right\} = \{2\cdot\bs{b}_3\,\ts{rad}\cdot\ts{s}^{-1},100^\circ\}$}, obtained using the switching controller. \textbf{(c)}~Evolutions of $\Lambda$ and $\sigma$ over time corresponding to the flight test with data in~(b), computed by the switching controller in real time. \textbf{(d)}~Reference and measured yaw angle, for the pair of initial conditions \mbox{$\left\{ \bs{\omega}_0, \psi_0 \right\} = \{4\cdot\bs{b}_3\,\ts{rad}\cdot\ts{s}^{-1},100^\circ\}$}, obtained using the benchmark controller. \textbf{(e)}~Reference and measured yaw angle, for the pair of initial conditions \mbox{$\left\{ \bs{\omega}_0, \psi_0 \right\} = \{4\cdot\bs{b}_3\,\ts{rad}\cdot\ts{s}^{-1},100^\circ\}$}, obtained using the switching controller. \textbf{(f)}~Evolutions of $\Lambda$ and $\sigma$ over time, corresponding to the flight test with data in~(e), computed by the switching controller in real time. \label{Fig02}}
\vspace{-2ex}
\end{figure*}

Now, using \textit{Lyapunov's indirect method}, we show that \mbox{$\{\bs{\qbar}^{\dagger}_{\ts{e},+1}, \bs{\nu}^{\star}_{+1} \}$} is an unstable saddle point. For this purpose, we first find the Jacobian matrix of the CL system specified by (\ref{EQ08}) with \mbox{$\sigma = +1$}. Specifically,
\begin{align}
\begin{split}
&\bs{A} \left( \bs{\qbarsmall}_{\ts{e}},\bs{\nu}_{+1} \right) =\\ 
& \hspace{-0.8ex}\left[
 \begin{matrix}
    0 & -\frac{1}{2}\bs{\nu}_{+1}^T+k_{\bs{n}}\bs{n}_{\ts{e}}^T & - \frac{1}{2}\bs{n}_{\ts{e}}^T\\
    \vspace{-2ex}
    \\
    \frac{1}{2}(\bs{\nu}_{+1} - k_{\bs{n}}\bs{n}_{\ts{e}}) & \frac{1}{2}\left([\bs{\nu}_{+1}]_{\times} - k_{\bs{n}}m_{\ts{e}}\bs{I}\right) &  \frac{1}{2}(- [\bs{n}_{\ts{e}}]_{\times} + m_{\ts{e}}\bs{I})\\
     \vspace{-2ex}
    \\
    \bs{0}&-k_{\bs{\qbarssmall}}\bs{I} & -k_{\bs{\omega}}\bs{I}
  \end{matrix}   
\right],
\end{split}
\label{EQ15}
\end{align}
where the terms $[\bs{\nu}_{+1}]_{\times}$ and $[\bs{n}_{\ts{e}}]_{\times}$ denote the \mbox{skew-symmetric} \mbox{cross-product} matrices for $\bs{\nu}_{+1}$ and $\bs{n}_{\ts{e}}$, as defined in~\cite{markley2014fundamentals}, respectively. Next, we evaluate the Jacobian matrix in~(\ref{EQ15}) at \mbox{$\{\bs{\qbar}^{\dagger}_{\ts{e},+1}, \bs{\nu}^{\star}_{+1} \}$}, which yields
\begin{align}
\bs{A} ( \bs{\qbarsmall}^{\dagger}_{\ts{e},+1}, \bs{\nu}^{\star}_{+1})= \hspace{-0.6ex}
\left[
\begin{array}{ccc}
0 & \bs{0} & \bs{0} \\
\vspace{-2ex}
\\
\bs{0} & \frac{1}{2}k_{\bs{n}}\bs{I} & -\frac{1}{2}\bs{I} \\
 \vspace{-2ex}
\\
\bs{0} & -k_{\bs{\qbarssmall}}\bs{I} & -k_{\bs{\omega}}\bs{I} 
\end{array}
\right]\hspace{-0.3ex}.
\label{EQ16}
\end{align}
Clearly, one of the eigenvalues of \mbox{$\bs{A} ( \bs{\qbarsmall}^{\dagger}_{\ts{e},+1}, \bs{\nu}^{\star}_{+1})$} is $0$. The other six eigenvalues can be found by analyzing the remaining \mbox{$6 \times 6$} submatrix after the first column and first row of the matrix are removed. Thus, it can be shown that \mbox{$\bs{A} ( \bs{\qbarsmall}^{\dagger}_{\ts{e},+1}, \bs{\nu}^{\star}_{+1})$} has two additional \textit{real} eigenvalues, each with algebraic multiplicity of~$3$. These two eigenvalues can be computed as
\begin{align}
\lambda_i = \frac{-k_{\bs{\omega}}+\frac{1}{2}k_{\bs{n}} \pm \left[\left(-k_{\bs{\omega}}+\frac{1}{2}k_{\bs{n}}\right)^2+ 2(k_{\bs{\qbars}}+k_{\bs{n}}k_{\bs{\omega}})\right]^{\frac{1}{2}}}{2},
\label{EQ17}
\end{align}
with \mbox{$i \in \left\{1,2\right\}$}. Furthermore, from simply observing that \mbox{$\left[\left(-k_{\bs{\omega}}+\frac{1}{2}k_{\bs{n}}\right)^2+ 2(k_{\bs{\qbars}}+k_{\bs{n}}k_{\bs{\omega}})\right]^{\frac{1}{2}}>-k_{\bs{\omega}}+\frac{1}{2}k_{\bs{n}}$,~~we~can} conclude that  \mbox{$0 < \lambda_1 \in \mathbb{R}$} and \mbox{$0 > \lambda_2 \in \mathbb{R}$}, which implies that \mbox{$\{\bs{\qbar}^{\dagger}_{\ts{e},+1}, \bs{\nu}^{\star}_{+1} \}$} is an unstable saddle point and, therefore, that there is at least one trajectory converging to this point. Also, by noting that \mbox{$V_{+1}(\bs{\qbar}^{\dagger}_{\ts{e}}, \bs{\nu}^{\star}_{+1}) = 4c$}, we provide an estimate for the region of attraction of the stable equilibrium point~\mbox{$\{\bs{\qbar}^{\star}_{\ts{e},+1}, \bs{\nu}^{\star}_{+1} \}$}, which is given by~\mbox{$\mathcal{D}_{+1} = \{\{ \bs{\qbar}_{\ts{e}},\bs{\nu}_{+1}\} \in \{\mathcal{S}^3,\mathbb{R}^3 \}~|~V_{+1}(\bs{\qbar}_{\ts{e}}, \bs{\nu}_{+1})<4c\}$. The} logical argument for this conclusion is as follows. For an appropriate choice of $c$, \mbox{$\dot{V}_{+1}(\bs{\qbar}_{\ts{e}},\bs{\nu}_{+1})<0$} and, as a consequence, if \mbox{$V_{+1}(\bs{\qbar}_{\ts{e}}, \bs{\nu}_{+1})<4c$}, \mbox{$V_{+1}(\bs{\qbar}_{\ts{e}}, \bs{\nu}_{+1})\rightarrow 0$} as \mbox{$t\rightarrow \infty$}. Therefore, all trajectories starting inside $\mathcal{D}_{+1}$ necessarily converge to \mbox{$\{\bs{\qbar}^{\star}_{\ts{e},+1}, \bs{\nu}^{\star}_{+1} \}$}. Note that if \mbox{$V_{+1}(\bs{\qbar}_{\ts{e}}, \bs{\nu}_{+1})>4c$}, no conclusion can be drawn using this line of thought because \mbox{$\{\bs{\qbar}^{\dagger}_{\ts{e},+1}, \bs{\nu}^{\star}_{+1} \}$} is an unstable saddle point. It is important to note that as $c$ increases, the region of attraction also increases; however, as shown below, this increase would directly affect the \mbox{Lyapunov-based} switching function of the CL system. A further result regarding the stability of the system can be obtained when \mbox{$c=1$}. In this case, (\ref{EQ13}) can be rewritten as
\begin{align}
\begin{split}
\dot{V}_{+1}(\bs{\qbar}_{\ts{e}},\bs{\nu}_{+1}) &= -k_{\bs{\qbars}}^{-1}k_{\bs{\omega}} \|\bs{\nu}_{+1}\|_2^2 - k_{\bs{n}} (1-m^2_{\ts{e}})\\
&\leq -\left[k_{\bs{\qbars}}^{-1}k_{\bs{\omega}} \|\bs{\nu}_{+1}\|_2^2 + k_{\bs{n}} (1-m_{\ts{e}})\right], \\
\end{split}
\label{EQ18}
\end{align}
for \mbox{$m_{\ts{e}} > 0$}. Thus, if we select \mbox{$k_{\bs{n}} = 4k_{\bs{\omega}}$}, it follows that 
\begin{align}
\begin{split}
\dot{V}_{+1}(\bs{\qbar}_{\ts{e}},\bs{\nu}_{+1}) \leq -2k_{\bs{\omega}}V_{+1}(\bs{\qbar}_{\ts{e}},\bs{\nu}_{+1}),
\end{split}
\label{EQ19}
\end{align}
which proves the \textit{exponential} stability of \mbox{$\{\bs{\qbar}^{\star}_{\ts{e},+1}, \bs{\nu}^{\star}_{+1} \}$} in \mbox{$\mathcal{D}_{\ts{e},+1} = \{\{ \bs{\qbar}_{\ts{e}},\bs{\nu}_{+1}\} \in \{\mathcal{S}^3,\mathbb{R}^3 \}~|~m_{\ts{e}} > 0\}$} by directly applying the proof of \mbox{Theorem\,$4.10$} in~\cite{khalil2002nonlinear}. An almost identical procedure can be used to study the stability properties of the subsystem corresponding to \mbox{$\sigma = -1$}. Specifically, following the approach in~\cite{Goncalves2024ROBOT}, it can be shown that \mbox{$\{\bs{\qbar}^{\dagger}_{\ts{e},-1}, \bs{\nu}^{\star}_{-1}\}$} is locally asymptotically stable, using the LF defined as 
\begin{align}
\begin{split}
V_{-1}(\bs{\qbar}_{\ts{e}}, \bs{\nu}_{-1}) = \frac{1}{2} k_{\bs{\qbars}}^{-1} \bs{\nu}_{-1}^T \bs{\nu}_{-1} + 2c(1 + m_\ts{e}),
\end{split}
\label{EQ20}
\end{align}
and that \mbox{$\{\bs{\qbar}^{\star}_{\ts{e},-1}, \bs{\nu}^{\star}_{-1}\}$} is unstable, using Lyapunov's indirect method. 

Last in this section, we define a switching function $\Lambda$ that selects $\sigma$ in a manner such that \mbox{Theorem\,$3.1$} in~\cite{liberzon2003switching} can be applied to guarantee the asymptotic stability of the CL switching dynamics. With this purpose, we define 
\begin{align}
\Delta V = V_{-1}  - V_{+1} = -2k_{\bs{\qbars}}^{-1}k_{\bs{n}}\bs{\omega}^T_\ts{e}\bs{n}_\ts{e} + 4cm_\ts{e},
\label{EQ21}
\end{align}
and select \mbox{$\Lambda = \Delta V$} to ensure that the instantaneous value of the switching system's LF always decreases when a switching event occurs. Namely, for a transition from \mbox{$\sigma = +1$} to $-1$, \mbox{$\Delta V \leq -\delta < 0$}, which implies that \mbox{$V_{-1} - V_{+1} < 0$}; and, for a transition from \mbox{$\sigma = -1$} to $+1$, \mbox{$\Delta V \geq \delta > 0$}, which implies that \mbox{$V_{+1} - V_{-1} < 0$}. In summary, the chosen switching condition guarantees that the instantaneous value of the switching system's LF always decreases---during continuous operation because \mbox{$\dot{V}_{\sigma} < 0$}, for \mbox{$\sigma \in \left\{+1,-1 \right\}$}, and across transitions because we set \mbox{$\Lambda = \Delta V$}. Formally,
\begin{align}
V_{\sigma(t_l)} \left( \bs{\qbar}_{\ts{e}}(t_l), \bs{\nu}_{\sigma(t_l)}(t_l) \right) - V_{\sigma(t_j)} \left( \bs{\qbar}_{\ts{e}}(t_j), \bs{\nu}_{\sigma(t_j)}(t_j) \right) \leq -\delta,
\label{EQ22}
\end{align}
where \mbox{$t_j < t_k < t_l$} and \mbox{$\sigma(t_j) = \sigma(t_l) \neq \sigma(t_k)$}, which reiterates the fact that when a switching occurs between two different instants, the value of each subsystem's LF decreases. Accordingly, since both subsystems specified by~(\ref{EQ08}) have locally asymptotically stable equilibrium points and, in addition, condition~(\ref{EQ22}) holds, the local asymptotic stability of the set \mbox{$\{\sigma \bs{\qbar}^{\star}_\ts{e}, \bs{\nu}^{\star}_{\sigma}\}$} can be directly inferred from \mbox{Theorem\,$3.1$} in~\cite{liberzon2003switching}. 

\hfill $\square$
\begin{table*}[t!]
\vspace{1ex}
\renewcommand{\arraystretch}{1.4}
\centering
\caption{Values of the Lyapunov function $V_{\sigma}$ for different pairs of tested initial conditions. \label{Table01}}
\vspace{-1ex}
\begin{tabular}{|c|c|c|c|c|c|}
\hline
\rowcolor{gray!30}
$\left\{\bs{\omega}_0, \psi_0\right\}$ & $\{2\cdot\bs{b}_3\,\ts{rad}\cdot\ts{s}^{-1},150^\circ\}$ & $\{3\cdot\bs{b}_3\,\ts{rad}\cdot\ts{s}^{-1},120^\circ\}$ & $\{4\cdot\bs{b}_3\,\ts{rad}\cdot\ts{s}^{-1},100^\circ\}$ & $\{2\cdot\bs{b}_3\,\ts{rad}\cdot\ts{s}^{-1},100^\circ\}$ & $\{2\cdot\bs{b}_3\,\ts{rad}\cdot\ts{s}^{-1},210^\circ\}$ \\ [0.3ex]
\hline 
\rowcolor{gray!10}  
$V_{\sigma}(t_0)$ & $7.97$ & $7.60$ & $7.24$ & $6.10$ & $5.90$\\
\hline
\end{tabular}
\vspace{-1ex}
\end{table*}
\section{Experimental Results} 
\label{Section05}
The flight tests discussed in this section were performed using the experimental setup described in~\cite{Goncalves2024ICRA,Goncalves2024ROBOT}. To test and demonstrate the suitability and performance of the proposed switching scheme---and compare the obtained experimental data to those obtained with the benchmark controller---we carried out one hundred experiments using the \mbox{$31$-g} \mbox{Crazyflie\,$2.1$} platform~\cite{bitcraze}, shown in \mbox{Fig.\,\ref{FIG01}}. Two sets of flight experiments are shown in Fig.\,\ref{Fig02}. During each experiment, the flyer tracks a \mbox{high-speed} \mbox{yaw-angle} reference composed of three stages, as indicated in~\mbox{Figs.\,\ref{Fig02}(a)}~and~(d). In \mbox{Stage\,$1$}, the UAV hovers while closely tracking a zero \mbox{yaw-angle} reference, i.e., \mbox{$\psi_{\ts{d}} = 0^\circ$}; in \mbox{Stage\,$2$}, the UAV rotates along the yaw axis, $\bs{b}_3$, by closely tracking a constant \mbox{angular-velocity} reference, $\bs{\omega}_0$, until reaching a predefined yaw angle, $\psi_0$, at time $t_0$, which is the initial time of \mbox{Stage\,$3$}; last, during \mbox{Stage\,$3$}, the \mbox{yaw-angle} reference $\psi_{\ts{d}}$ is set back to $0^\circ$, which compels the UAV to rotate toward it. In the cases presented here, the instant of the transition from \mbox{Stage\,$2$} to \mbox{Stage\,$3$} is when the switching controller selects the more \mbox{energy-efficient} torque input, which depends on the instantaneous AEQ and \mbox{angular-velocity} error, according to the logic discussed in Section\,\ref{Section04A}. Note that, during flight, the switching control algorithm continually decides if a transition between the two torque inputs specified by (\ref{EQ06}) is necessary; however, only abrupt changes in the attitude reference prompt the flight controller to switch between control torques. In the experimental cases presented in this section, we heuristically selected the gains for the benchmark controller to be \mbox{$k_{\bs{\qbars}} = 1000~\ts{N} \cdot \ts{m}$} and \mbox{$k_{\bs{\omega}} = 100~\ts{N} \cdot  \ts{m} \cdot \ts{s} \cdot \ts{rad}^{-1}$}; for the proposed switching control scheme, we selected the gains to be \mbox{$k_{\bs{\qbars}} = 10~\ts{N} \cdot \ts{m}$}, \mbox{$k_{\bs{\omega}} = 100~\ts{N} \cdot  \ts{m} \cdot \ts{s} \cdot \ts{rad}^{-1}$}, and \mbox{$k_{\bs{n}} = 10~\ts{rad} \cdot \ts{s}^{-1}$}. We chose these parameter values in order to implement the laws specified by both (\ref{EQ05})~and~(\ref{EQ06}) with almost identical proportional and derivative gains---note that the expression that defines $\bs{\nu}_{\sigma}$ contains an additional proportional term---and thus make the performance comparison between both tested controllers as meaningful as possible. Also, we selected \mbox{$c=2$}, which ensures that all the tested pairs of initial conditions at time $t_0$, $\left\{\bs{\omega}_0,\psi_0 \right\}$, define a state inside the region of attraction~\mbox{$\mathcal{D}_{\sigma} = \{\{ \bs{\qbar}_{\ts{e}},\bs{\nu}_{\sigma}\} \in \{\mathcal{S}^3,\mathbb{R}^3 \}~|~V_{\sigma}(\bs{\qbar}_{\ts{e}}, \bs{\nu}_{\sigma})<8\}$}, as seen in~\mbox{Table\,\ref{Table01}}.

\mbox{Figs.\,\ref{Fig02}(a)~and~(b)} show the desired and measured yaw angles, $\psi_{\ts{d}}$ and $\psi$, corresponding to two flight tests with the pair of initial conditions \mbox{$\left\{ \bs{\omega}_0, \psi_0 \right\} = 
\left\{2\cdot\bs{b}_3\,\ts{rad}\cdot\ts{s}^{-1},100^\circ \right\}$}, employing the benchmark and switching controller, respectively. \mbox{Fig.\,\ref{Fig02}(c)} shows the \mbox{time-evolutions} of the switching function, $\Lambda$, and switching signal, $\sigma$, computed in real time by the switching controller during the flight test corresponding to \mbox{Fig.\,\ref{Fig02}(b)}. \mbox{Figs.\,\ref{Fig02}(d)~and~(e)} show the desired and measured yaw angles, $\psi_{\ts{d}}$ and $\psi$, corresponding to two flight tests with the pair of initial conditions \mbox{$\left\{ \bs{\omega}_0, \psi_0 \right\} = 
\left\{4\cdot\bs{b}_3\,\ts{rad}\cdot\ts{s}^{-1},100^\circ \right\}$}, employing the benchmark and switching controller, respectively. \mbox{Fig.\,\ref{Fig02}(f)} shows the \mbox{time-evolutions} of the switching function, $\Lambda$, and switching signal, $\sigma$, computed in real time by the switching controller during the flight test corresponding to \mbox{Fig.\,\ref{Fig02}(e)}. For the first pair of initial conditions, corresponding to an \mbox{angular-velocity} magnitude of \mbox{$\|\bs{\omega}_0\|_2 = 2\,\ts{rad}\cdot\ts{s}^{-1}$}, the abrupt change in the attitude reference induces a decrease in the value of $\Lambda$ that is not sufficiently large to activate a switching event. In contrast, for the second pair of initial conditions, corresponding to an \mbox{angular-velocity} magnitude of \mbox{$\|\bs{\omega}_0\|_2 = 4\,\ts{rad}\cdot\ts{s}^{-1}$}, the abrupt change in the attitude reference induces a decrease in the value of $\Lambda$ that exceeds the hysteresis margin $\delta$ and, consequently, causes a transition from \mbox{$\sigma = +1$} to $-1$. Video footage of these and other relevant flight tests can be viewed in the accompanying supplementary movie.
\begin{figure}[t!]
\vspace{1ex}
\begin{center}
\includegraphics[width=0.48\textwidth]{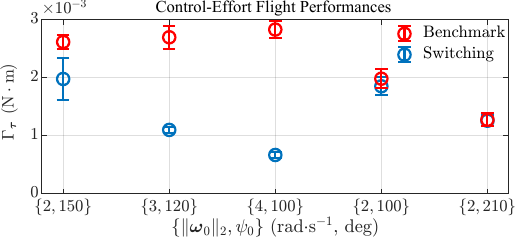}
\end{center}
\vspace{-3ex}
\caption{\textbf{Comparison of achieved flight performances corresponding to five different pairs of initial conditions and two tested controllers (benchmark and switching).} Here, each data point shows the mean and ESD of the values of $\Gamma_{\bs{\tau}}$ corresponding to ten \mbox{back-to-back} flight experiments.
\label{Fig03}}
\vspace{-2ex}
\end{figure}

To assess and compare the flight performances achieved with the switching and the benchmark controllers, we use the PFM defined as 
\begin{align}
\vspace{-1ex}
\Gamma_{\bs{\tau}} = \left(\frac{1}{t_{\ts{f}} - t_{0}}\int^{t_\ts{f}}_{t_{0}} \|\bs{\tau}_{\hspace{-0.2ex}\ts{exp}}(t)\|_2^2 dt\right)^{\frac{1}{2}}, ~~~~[\ts{N}\cdot \ts{m}],~~~~~~~
\label{EQ23}
\end{align}
where \mbox{$t_{\ts{f}} = t_0 + 3$\,s}; and, \mbox{$\bs{\tau}_{\hspace{-0.2ex}\ts{exp}}\in\left\{\bs{\tau}_{\ts{b}},\bs{\tau}_{\hspace{-0.2ex}\sigma}\right\}$} is the control law used during a given experiment. \mbox{Fig.\,\ref{Fig03}} presents \mbox{empirically-estimated} flight performances for five different pairs of initial conditions, computed from data obtained with the switching and benchmark controllers. For each set of initial conditions and controller, we show the mean and \textit{empirical standard deviation} (ESD) of the values of $\Gamma_{\bs{\tau}}$ corresponding to ten \mbox{back-to-back} flight experiments. The first three pairs of initial conditions correspond to cases in which the benchmark and switching controllers select a different direction for the application of the torque input; namely, \mbox{$\ts{sgn}\left\{m_{\ts{e}}\right\} \neq \sigma$}. In these examples, the switching controller reduces the control effort, $\Gamma_{\bs{\tau}}$, by about $53\,\%$, on average, with respect to that corresponding to the benchmark controller. Notably, for the same pair of initial conditions, the worst performance achieved with the switching controller is always better than the best performance achieved with the benchmark controller (see \mbox{Fig.\,\ref{Fig03}}). The last two pairs of initial conditions correspond to cases in which the switching and benchmark controllers select the same direction for the torque input; namely, \mbox{$\ts{sgn}\left\{m_{\ts{e}}\right\} = \sigma$}. As expected, in these cases, the means and ESDs of the values of $\Gamma_{\bs{\tau}}$ corresponding to both controllers are very similar, which is consistent with \mbox{bare-eye} observations of the UAV during flight.

\section{Conclusions}
\label{Section06}
We presented a method for synthesizing and implementing a new type of switching attitude controllers for UAVs that account for both attitude and \mbox{angular-velocity} errors when computing the torque control inputted to the system during flight. For the purposes of controller design and stability analysis, we employed classical nonlinear theory and contemporary dynamic switching techniques. In particular, we introduced a new type of LF that enabled us to compute a larger estimate of the CL system's region of convergence, compared to those estimated in previous works. Furthermore, this approach provided us with the conditions required to guarantee exponential stability. To test and demonstrate the functionality, suitability, and performance of the proposed approach, we performed one hundred \mbox{high-speed} \mbox{yaw-tracking} flight tests using a \mbox{$31$-g} quadrotor flown with the proposed switching controller and a \mbox{commonly~used} benchmark switching scheme. Simultaneously, we also empirically demonstrated that all the pairs of initial conditions used in the flight tests lie inside the estimated region of attraction of the CL system. The experimental data obtained through the flight tests show that the control effort is reduced by about $53\,\%$, on average, when the switching controller---in contrast with the case of the benchmark scheme---decides not to follow the shorter rotational path to eliminate the attitude error after the attitude reference is changed abruptly.  
 
\balance
\bibliographystyle{IEEEtran}
\bibliography{paper}
\end{document}